\def\red#1{{
#1
}}
\def\gre#1{{
#1
}}
\def\blu#1{{
#1
}}

\def\beq{\begin{equation}}
\def\eeq{\end{equation}}
\def\beqn{\begin{eqnarray}}
\def\eeqn{\end{eqnarray}}

\def\inbar{\,\vrule height1.5ex width.4pt depth0pt}

\def\IC{\relax\hbox{$\inbar\kern-.3em{\rm C}$}}
\def\IQ{\relax\hbox{$\inbar\kern-.3em{\rm Q}$}}
\def\IR{\relax{\rm I\kern-.18em R}}
 \font\cmss=cmss10 \font\cmsss=cmss10 at 7pt
\def\IZ{\relax\ifmmode\mathchoice
 {\hbox{\cmss Z\kern-.4em Z}}{\hbox{\cmss Z\kern-.4em Z}}
 {\lower.9pt\hbox{\cmsss Z\kern-.4em Z}}
 {\lower1.2pt\hbox{\cmsss Z\kern-.4em Z}}\else{\cmss Z\kern-.4em Z}\fi}

\def\AEF{Faraggi A E }
\def\NPB#1#2#3{ 19#2  {\it Nucl.\ Phys.\ B}\/ {\bf #1}  #3}
\def\PLB#1#2#3{ 19#2  {\it Phys.\ Lett.\ B}\/ {\bf #1}  #3}
\def\PLA#1#2#3{ 19#2  {\it Phys.\ Lett.\ A}\/ {\bf #1}  #3}
\def\PRD#1#2#3{ 19#2  {\it Phys.\ Rev.\ D}\/ {\bf #1}  #3}

\def\PRT#1#2#3{  19#2  {\it Phys.\ Rep.}\/ {\bf#1}  #3}
\def\MODA#1#2#3{ 19#2  {\it Mod.\ Phys.\ Lett.\ A}\/ {\bf #1}  #3}
\def\IJMP#1#2#3{ 19#2  {\it Int.\ J.\ Mod.\ Phys.\ A}\/ {\bf #1} #3}

\font\bigbf=cmssbx10 scaled\magstep2

\font\it=cmti10 at 12pt
\font\ss=cmss10 at 12pt

\ss                                     
\documentstyle[iopconf1]{article}
\begin{document}
\rightline{UMN--TH--1819--99}
\rightline{TPI--MINN--99/42}
\rightline{hep-ph/9910042}
\rightline{October 1999}
\title{~~~~~~~~~~~- Superstring Phenomenology -\\
~~~~~~~~~~~~~~~~A Personal Perspective\footnote{Invited talk presented 
at Beyond the Desert 99, Castle Ringberg,
Tegernsee, Germany, 6-12 June 1999.}
}
\author{Alon E. Faraggi
\dag\footnote{E-mail: faraggi@mnhepo.hep.umn.edu.}
}
\affil{\dag\ Theoretical Physics Institute, 
Department of Physics, University of Minnesota, 
Minneapolis, MN 55455}
\beginabstract
In the first part of this paper I review the construction
of the realistic free fermionic models,
as well as current attempts to study aspects of these models
in the nonperturbative framework of M-- and F--theories.
I discuss the recent demonstration of a Minimal Superstring
Standard Model, which contains in the observable sector,
below the string scale, solely the MSSM charged spectrum,
and provides further support to the assertion that the true
string vacuum is connected to the $Z_2\times Z_2$ orbifold
in the vicinity of the free fermionic point in the Narain moduli
space. In the second part I review the recent formulation of
quantum mechanics from an equivalence postulate, which offers
a new perspective on the synthesis of gravity and quantum
mechanics, and contemplate possible relations with string theory
and beyond. 
\endabstract
\section{Introduction}
Superstring phenomenology aims at achieving two goals.
The first task is to reproduce the phenomenological
data provided by the Standard Particle Model. The
subsequent goal is to extract possible experimental
signatures which may provide further evidence for 
the validity of specific string models, in particular,
and for string theory, in general. One should however
note that, in the lack of substantial experimental
evidence for any extension of the Standard Model, the
conservative approach would be to derive solely the
Standard Model, which we may assume to include 
non--vanishing neutrino masses. Experimental
signatures beyond the Standard Model become
firm theoretical predictions once the first
goal is achieved and it appears that something
extra {\it unavoidably} remains.

Despite its experimental success, the Standard
Model leaves much to be desired. In the first place
the Standard Model is made of several disparate sectors.
These include the matter, the interaction, and the Higgs, sectors.
The Higgs sector is
still unobserved experimentally and the least understood.
The matter and interaction sectors are made of
similar but distinct elements, like the different gauge groups
of each interaction and the multiplicity of generations,
which are parametrized by various parameters. This enumeration
is clearly unappealing and it is reasonable to seek a more
economical description. The most important guide in this quest
is the multiplet structure of the Standard Model, which is
exhibited below,

${~~~~~~SU(3)_C}\times {SU(2)_L}\times {U(1)_Y}~~\longrightarrow~~%
{SU(5)}~,%
~~~~~~~~~~~~{SO(10)}~,~~~~~~~{E_6}$

$Q~~:~(~{3}~~~~~,~~~~~{2}~~~~~,~~~~{{1\over6}}~)$

$U_L^c~:~(~{{\bar3}}~~~~~,~~~~~{1}~~~~~,~{-{2\over3}}~)$~~~
~~$\rightarrow$~~~~~${10}$

$E_L^c~:~(~{1}~~~~~,~~~~~{1}~~~~~,~{+1}~)$

~~~~~~~~~~~~~~~~~~~~~~~~~~~~~~~~~~~~~~~~~~~~~~~~~~~~~~~~~~~~~~~~~~~~~~~~%
$\rightarrow$~~~~${16}$~~~~~~$\rightarrow$~~~~${27}$

$L~~~:~(~{1}~~~~~,~~~~~{2}~~~~~,~{-{1\over2}}~)$

$D_L^c~:~(~{{\bar3}}~~~~~,~~~~~{1}~~~~~,
~~~{{1\over3}}~)$~~~~~~~$\rightarrow$~~~~~${{\bar5}}$

~~~~~~~~~~~~~~~~~~~~~~~~~~~~~~~~~~~~~~~~~~~~~~~~~~

$N_L^c~:~(~{1}~~~~~,~~~~~{1}~~~~~,~~~{0}~)$~~~
~~~~~~~~~~~~~~~~$+~~{1}$

The matter and gauge multiplets of the Standard Model
amazingly fit into representations of larger unifying
gauge groups \cite{gg}. Most appealing is the framework
of $SO(10)$, in which all the Standard Model
states (including the right--handed neutrinos
which are desirable for neutrino masses and
oscillations), in each generation, are embedded
in a single representation. A priori there was no reason
for this to have been the case. But strikingly
all three generations fit, each, into a fundamental 
representation of $SO(10)$. 
If we regard (as we should) the quantum numbers
of the Standard Model states as experimental observables,
then this scheme correlates 18 observable parameters.
It seems to me therefore that to deny the evidence for the underlying
$SO(10)$ structure of the Standard Model is synonymous
to dismissing the Standard Model itself. 

An important experimental fact is the 
unobservation of proton decay. 
In the Standard Model the proton decay is
forbidden by renormalizability and accidental global
symmetries. In general, extensions of the Standard Model
produce proton decay mediating operators. 
In non--SUSY GUTs proton decay is mediated by dimension
six operators. In supersymmetric theories dimension four
and five operators are also generically allowed. Proton decay
becomes an especially acute problem when gravity
is unified with the gauge interactions because
in that case renormalizability and global symmetries are not expected
to be respected. In this context therefore it is
expected that proton stability can only be maintained
if there exist a gauge symmetry, which after its breaking
still leaves a residual discrete symmetry, 
which forbids proton decay \cite{lds}.
This is very hard and nontrivial to achieve. The evidence
for unification, provided by the Standard Model multiplet
structure, together with proton longevity, indicate that
the Standard Model cannot be strongly perturbed up to
a very large scale. It is difficult to envision
how a strong perturbation of the Standard Model
at a low scale will not run into conflict with
the proton lifetime. These experimental facts therefore
indicate the big desert scenario. 

Unification and the big desert scenario is then supported
by another observation. Namely, if one extrapolates
the Standard Model gauge couplings, they are
seen to converge at a high scale, which is
one or two orders of magnitude below the heterotic
string scale \cite{gqw}. This picture is especially appealing 
in the case of supersymmetric theories, where
the couplings are seen to meet at a scale which is
of the order of $10^{16}{\rm GeV}$ \cite{gcumssm}.
This extrapolation should be taken as qualitative
support for the consistency of the big desert scenario.
The appealing feature of supersymmetric theories
is the fact that when the symmetry is local it necessitates
the appearance of a spin two field. We then see that
the gauge and gravitational interactions start to
converge into a unifying setting. Furthermore, this
setting also provides the means to understand how
a very small scale such as the electroweak scale
can be generated by extrapolation from the
Planck scale. 

Despite their enormous success point quantum 
field theories still leave many questions unresolved.
Why is a particular gauge group observed at low energies,
together with the multiplicity of generations ?
The proliferation of Standard Model parameters,
in particular in the flavor sector, and the hierarchy
between them does not have its origin in GUTs or
SUSY GUTs. To understand these issues we must
incorporate gravity into the picture. Most
importantly, point quantum field theories
do not provide for a consistent formulation of
quantum gravity. A consistent formulation of
quantum gravity requires new conceptual framework
and tools. Such a framework will then also
shed light on the structure of the Standard Model.

String theory provides a consistent
perturbative formulation of quantum gravity. 
String theory is unique in the sense
that it is the only approach to date which
gives a consistent common framework for both gravity and
the gauge interactions. As such string theory
exactly suits our purpose, {\it i.e.} it provides
the tools to study how the Standard
Model structure and parameters may arise from a
theory of quantum gravity. 

String theory is defined in perturbation theory.
As such it is clear that string theory cannot be 
the final story. Indeed, over the last few years
an important new understanding has emerged in which
it is seen that all the different string theories
in ten dimensions are in fact perturbative limits
of a single theory. This is a very encouraging
picture because it tells us that by utilizing string
perturbation theory we are truly probing the 
underlying nonperturbative theory. Now suppose
that the situation was reversed and we first
had in our hands the full nonperturbative formulation. 
It is likely that in that case what we would have done
in order to study its connection with the real 
world is to develop perturbation theory in
the vicinity of its most relevant limits.
\section{Superstring constructions}
There are two complementary approaches to superstring
phenomenology. In one, the general strategy is to first try to
understand what is the nonperturbative formulation of string theory.
The hope is that the unique string vacuum will be fixed
and the low energy predictions unambiguously determined. The second,
asserts that we must use low energy data to single
out phenomenologically interesting superstring vacua. Such string 
models will then be instrumental to understand the dynamics
which select the string vacuum. These two approaches are in a sense
complementary and progress is likely to be made by pursuing
both approaches in parallel.

The general goal is therefore to construct superstring models
that are as realistic as possible. A realistic model of unification
must satisfy a large number of constraints,

\centerline{{$\underline{{\hbox{~~~~~~~~~~~~~~~~~~~~~~~~~~~~~~~~~~~~~}}}$}}
{}~~~~~$1.$ Gauge group ~$\longrightarrow$~$SU(3)\times SU(2)\times U(1)_Y$
~~~~~~~~~~~~~~~{$U(1)\in SO(10)$}

{}~~~~~$2.$ Contains three generations

{}~~~~~$3.$ Proton stable ~~~~~~~~($\tau_{\rm P}>10^{30+}$ years)

{}~~~~~$4.$ N=1 supersymmetry~~~~~~~~(or N=0)

{}~~~~~$5.$ Contains Higgs doublets $\oplus$ potentially realistic
Yukawa couplings

{}~~~~~$6.$ Agreement with $\underline{\sin^2\theta_W}$ and
$\underline{\alpha_s}$ at $M_Z$ (+ other observables).

{}~~~~~$7.$ Light left--handed neutrinos

{}~~~~~~~~~~~~~$~8.$ $SU(2)\times U(1)$ breaking

{}~~~~~~~~~~~~~$~9.$ SUSY breaking

{}~~~~~~~~~~~~~$10.$ No flavor changing neutral currents

{}~~~~~~~~~~~~~$11.$ No strong CP violation

{}~~~~~~~~~~~~~$12.$ Exist family mixing and weak CP violation

{}~~~~~$13.$ +~~ {\bf ...}

{{}~~~~~$14.$ +~~~~~~~~~~~~~~~~{\bigbf{No free exotics}}}

It is important to emphasize that the
$SO(10)$ structure, advocated above,
need not be realized in an effective
field theory but can be broken directly at the string level.
In which case the Standard Model spectrum still
arises from $SO(10)$ representations, but the $SO(10)$
non--Abelian spectrum, beyond the Standard Model, 
is projected out by the GSO projections. However, 
if we take the Standard Model $SO(10)$ embedding
as a necessary requirement this means that the
weak hypercharge must have the standard $SO(10)$
embedding with $k_Y=5/3$.
This requirement then already excludes many of
the semi-realistic models, which have been constructed
to date. Similarly, the requirement that no
free exotic particles with fractional electric
charge remain in the massless spectrum imposes
a highly non--trivial constraint on otherwise
valid models. The phenomenological constraints
impose very restrictive constraints on the superstring
constructions. This is augmented by the fact
that, unlike in field theory model building,
in string model building both the entire spectrum and
symmetries are fixed in a given vacuum. One does
not have the freedom to add an additional $U(1)$ or
discrete symmetry, or additional matter, to suit one needs.  
Therefore, string model building is more restrictive
than field theory model building. A string model
that can satisfy all of the above requirements
is likely to be more than an accident. 

There are several possible ways to try to construct
realistic superstring models. One possibility is to
construct superstring models with an intermediate 
GUT, or semi--GUT gauge group, like $SU(5)$, $SO(10)$,
$E_6$, etc \cite{stringguts},
or $SU(3)^3$ \cite{suthree}, $SU(5)\times U(1)$
\cite{revamp} or $SO(6)\times SO(4)$ \cite{patisalamstrings},
which are broken to the Standard Model gauge group at
an intermediate energy scale.
The other possibility is to construct superstring models in which
the non--Abelian factors of the Standard--Model gauge group are obtained
directly at the string level \cite{zthree,fny,eu,top,otherrsm}.
The advantage in the second case, as well as in the Pati--Salam
type models, is that in these cases the
color Higgs triplets, which mediate proton decay through
dimension five operators, can be projected out by the GSO
projections. Such models then provide a superstring solution
to the GUT hierarchy problem \cite{ps}.

With the advent of superstring duality arguments we
can use the different perturbative string limits
to try to construct realistic string models. These
are all supposedly connected by duality relations
and a model in one limit should have a dual model
in another limit. Interesting alternatives
to the heterotic string are the type I constructions,
which allow the unification scale to be lowered as
the gauge and gravity multiplets in this case do not
arise from the same sector. However, the heterotic 
string framework still remains the most appealing
as it is the only one which naturally gives rise
to $SO(10)$ multiplets in the 16 representation.

The construction of realistic superstring vacua proceeds by studying
compactification of the heterotic string from ten to four dimensions. 
Various methods can be used for this purpose which include
geometric and algebraic tools, and each has its own advantages and
disadvantages. One class of models utilizes compactifications
on Calabi--Yau 3--folds that give rise to an $E_6$ observable
gauge group, which is broken further by Wilson lines
to $SU(3)^3$ \cite{suthree}. This type of geometrical 
compactifications correspond at special points to
conformal theories which have $(2,2)$ world--sheet supersymmetry.
Similar compactifications which have only (2,0) world--sheet
supersymmetry have also been studied and can lead to compactifications
with $SO(10)$ and $SU(5)$ observable gauge groups \cite{twozero}.
The analysis of this type of compactification
is complicated due to the fact that they do not correspond
to free world--sheet theories. Therefore, it is difficult
to calculate the parameters of the Standard Model in these constructions.
On the other hand they provide a sophisticated mathematical
window to the underlying geometry.

The next class of superstring vacua are the orbifold models \cite{dhvw}.
Here one starts with a compactification
of the heterotic string on a flat torus,
using the Narain prescription \cite{narain},
and utilizes free world--sheet bosons. The Narain
lattice is moded out by some discrete symmetries which are the
orbifold twisting. An important class of models of this type are
the $Z_3$ orbifold models \cite{zthree}. These give rise to
three generation models with $SU(3)\times SU(2)\times U(1)^n$
gauge group. A deficiency of this class of models is that
they do not give rise to the standard $SO(10)$ embedding of
the Standard Model spectrum. Consequently,
the normalization of $U(1)_Y$, relative to the non--Abelian
currents, is larger than 5/3, the standard $SO(10)$ normalization.
This results generically in disagreement with the observed low energy
values for $\sin^2\theta_W(M_Z)$ and $\alpha_s(M_Z)$. 

A special type of string compactifications that has been
studied in detail are the free fermionic models. The simplest
examples correspond to $Z_2\times Z_2$ orbifolds at special points
in the compactification space. These models give rise to the most
realistic superstring models constructed to date. They
produce three generation models with the standard $SO(10)$
embedding of the Standard Model spectrum. Hence in these
models $U(1)_Y$ has the standard $SO(10)$ embedding, with
$k_Y=5/3$. Consequently these models can be
in agreement with the observed low energy values for 
$\sin^2\theta_W(M)$ and $\alpha_s(M_Z)$. 
There are several key features of these models which
suggest that the true string vacuum is in the vicinity of these models.
First is the fact that the free fermionic models are formulated at a
highly symmetric point in the string compactification space.
The second is that the emergence of three generations is
correlated with the underlying structure of the
$Z_2\times Z_2$ orbifold compactification \cite{foc}.
Each of the Standard Model generations is obtained from
one of the twisted sectors and carries horizontal charges under
one of the orthogonal planes of the $Z_2\times Z_2$ orbifold.
These models then give a reason for the existence of three generation
in nature, as originating from the structure of the underlying geometry.
\section{Free fermionic models}
A model in the free fermionic
formulation \cite{FFF} is defined by a set of  boundary
condition basis vectors, and one--loop GSO phases, which are
constrained by the string consistency requirements,
and completely determine the vacuum structure of the models.
The physical spectrum is obtained by applying the generalized 
GSO projections. The Yukawa couplings 
and higher order nonrenormalizable terms in the superpotential
are obtained by calculating correlators between vertex
operators \cite{KLN}. 
The realistic free fermionic
models produce an ``anomalous'' $U(1)$ symmetry,
which generates a Fayet--Iliopoulos D--term \cite{dsw},
and breaks supersymmetry at the Planck scale. Supersymmetry
is restored by assigning non vanishing VEVs to a set of Standard
Model singlets in the massless string spectrum along flat F and D
directions. In this process nonrenormalizable terms,
$$\langle{V_1^fV_2^fV_3^b\cdot\cdot\cdot\cdot V_N^b\rangle},$$
become renormalizable operators, 
$$V_1^fV_2^fV_3^b{{\langle V_4^b\cdots V_N^b\rangle}/{M^{N-3}}}$$
in the effective low energy field theory.

The first five basis vectors of the realistic free fermionic
models consist of the NAHE set \cite{nahe,eu,foc}.
The gauge group after the NAHE set is
$SO(10)\times E_8\times SO(6)^3$ with $N=1$ space--time supersymmetry, 
and 48 spinorial $16$ of $SO(10)$, sixteen from each sector $b_1$,
$b_2$ and $b_3$. The three sectors $b_1$, $b_2$ and $b_3$ are
the three twisted sectors of the corresponding $Z_2\times Z_2$
orbifold compactification. The $Z_2\times Z_2$ orbifold is special
precisely because of the existence of three twisted sectors,
with a permutation symmetry with respect to the horizontal $SO(6)^3$
symmetries. The NAHE set is depicted in the table below which
highlights its cyclic permutation symmetry.
\vspace{0.15cm}

~~~~~~~~~~~~~~~~~~~~~~~~~~~~~~~~{\red{\bf THE NAHE SET}}
{\large
{
\beqn
 &&\begin{tabular}{c|c|ccc|c|ccc|c}
 ~ & $\psi^\mu$ & $\red{\chi^{12}}$ & $\blu{\chi^{34}}$ & $\gre{\chi^{56}}$ &
        $\bar{\psi}^{1,...,5} $ &
        \red{$\bar{\eta}^1$}&
        \blu{$\bar{\eta}^2$}&
        \gre{$\bar{\eta}^3$}&
        $\bar{\phi}^{1,...,8} $ \\
\hline
\hline
      {\bf 1} &  1 & 1&1&1 & 1,...,1 & 1 & 1 & 1 & 1,...,1 \\
          $S$ &  1 & \red{1}&\blu{1}&\gre{1} & 0,...,0 & 0 & 0 & 0 & 0,...,0 \\
\hline
  \red{${b}_1$} &  1 & \red{1}&0&0 & 1,...,1 & \red{1} & 0 & 0 & 0,...,0 \\
  \blu{${b}_2$} &  1 & 0&\blu{1}&0 & 1,...,1 & 0 & \blu{1} & 0 & 0,...,0 \\
  \gre{${b}_3$} &  1 & 0&0&\gre{1} & 1,...,1 & 0 & 0 & \gre{1} & 0,...,0 \\
\end{tabular}
   \nonumber\\
   ~  &&  ~ \nonumber\\
     &&\begin{tabular}{c|cc|cc|cc}
 ~&     \red{$y^{3,...,6}$}  &
        \red{${\bar y}^{3,...,6}$}  &
        \blu{$y^{1,2},\omega^{5,6}$}  &
        \blu{${\bar y}^{1,2},\bar{\omega}^{5,6}$}  &
        \gre{$\omega^{1,...,4}$}  &
        \gre{$\bar{\omega}^{1,...,4}$}   \\
\hline
\hline
    {\bf 1} & 1,...,1 & 1,...,1 & 1,...,1 & 1,...,1 & 1,...,1 & 1,...,1 \\
    $S$     & 0,...,0 & 0,...,0 & 0,...,0 & 0,...,0 & 0,...,0 & 0,...,0 \\
\hline
\red{${b}_1$} & \red{1,...,1} & \red{1,...,1} & 0,...,0 & 0,...,0 & 
                                          0,...,0 & 0,...,0 \\
\blu{${b}_2$} & 0,...,0 & 0,...,0 & \blu{1,...,1} & \blu{1,...,1} & 
                                          0,...,0 & 0,...,0 \\
\gre{${b}_3$} & 0,...,0 & 0,...,0 & 0,...,0 & 0,...,0 & 
                                     \gre{1,...,1} & \gre{1,...,1} \\
\end{tabular}
\nonumber
\eeqn
}}

The NAHE set is common to a large class of three generation
free fermionic models. The construction proceeds by adding to the
NAHE set three additional boundary condition basis vectors
which break $SO(10)$ to one of its subgroups, $SU(5)\times U(1)$,
$SO(6)\times SO(4)$ or $SU(3)\times SU(2)\times U(1)^2$,
and at the same time reduces the number of generations to
three, one from each of the sectors $b_1$, $b_2$ and $b_3$.
The various three generation models differ in their
detailed phenomenological properties. However, many of
their characteristics can be traced back to the underlying
NAHE set structure. One such important property to note
is the fact that as the the generations are obtained
from the three twisted sectors $b_1$, $b_2$ and $b_3$,
they automatically possess the Standard $SO(10)$ embedding.
Consequently the weak hypercharge, which arises as
the usual combination $U(1)_Y=1/2 U(1)_{B-L}+ U(1)_{T_{3_R}}$,
has the standard $SO(10)$ embedding. To date, of the
orbifold models that have been constructed, only the
free fermionic models have yielded such a structure. 

The massless spectrum of the realistic free fermionic models
then generically contains three generations from the
three twisted sectors $b_1$, $b_2$ and $b_3$, which are
charged under the horizontal symmetries. The Higgs spectrum
consists of three pairs of electroweak doublets from the 
Neveu--Schwarz sector plus possibly additional one or
two pairs from a combination of the two basis vectors
which extend the NAHE set. Additionally the models 
contain a number of $SO(10)$ singlets which are
charged under the horizontal symmetries and
a number of exotic states. 

Exotic states
arise from the basis vectors which extend the NAHE
set and break the $SO(10)$ symmetry. Consequently, they
carry either fractional $U(1)_Y$ or $U(1)_{Z^\prime}$ charge.
Such states are generic in superstring models
and impose severe constraints on their validity.
In some cases the exotic fractionally charged
states cannot decouple from the massless
spectrum, and their presence invalidates otherwise
viable models. In the NAHE based models the fractionally
charged states always appear in vector--like
representations. Therefore, in general mass
terms are generated from renormalizable or nonrenormalizable
operators. However, the mass terms which arise
from non--renormalizable terms will in general be suppressed,
in which case the fractionally charged states may have
intermediate scale masses. Here
I describe the analysis of a model in which
all the fractionally charged states decouple from
the massless spectrum at the cubic level of the superpotential 
and receive mass of the order of the string scale.
\section{Minimal Superstring Standard Model}
The superstring model under consideration \cite{fny}
is a typical three generation free fermionic model. It is
generated by the NAHE--set plus three additional
basis vectors $\{b_4,\alpha,\beta\}$, where $b_4$ preserves
the $SO(10)$ symmetry, $\alpha$ breaks $SO(10)~\rightarrow~
SO(6)\times SO(4)$ and $\beta$ breaks $SO(6)\times SO(4)~\rightarrow~
SU(3)\times SU(2)\times U(1)^2$. The massless spectrum consist of
three generations from the sectors $b_1$, $b_2$ and $b_3$, 
The Neveu--Schwarz (NS) sector produces, the gravity and gauge multiplets,
three pairs of  electroweak doublets 
$\{h_1, h_2, h_3, {\bar h}_1, {\bar h}_2, {\bar h}_3\}$,
seven pairs of $SO(10)$ singlets with observable $U(1)$ charges, 
$\{\phi_{12},{\bar\phi}_{12},
   \phi_{23},{\bar\phi}_{23},
   \phi_{13},{\bar\phi}_{13},
   \phi_{56},{\bar\phi}_{56},
   \phi_{56}^\prime,{\bar\phi}_{56}^\prime,
   \phi_{4},{\bar\phi}_{4},
   \phi_{4}^\prime,{\bar\phi}_{4}^\prime\}$,
and three scalars that are singlets
of the entire four dimensional gauge group, $\phi_1,\phi_2,\phi_3$.
The states from the NS sector and the sectors $b_1$, $b_2$ and $b_3$
are the only ones that transform solely under the observable,
$SU(3)_C\times SU(2)_L\times U(1)_{B-L}\times
U(1)_{T_{3_R}}\times U(1)_{1,\cdots,6}$
gauge group. The sectors $(I)+b_{1,2,3,4}+2\beta$ produce $SO(10)$
singlet matter states in the 16 vector representation of the hidden
$SO(16)$ gauge group, decomposed under the final hidden group.
The sectors with some combination of
$\{{\bf 1},b_1,b_2,b_3,b_4,\alpha\}$ plus $\beta$ or $2\beta$
produce states that are $SU(3)_C\times SU(2)_L$
singlets, but carry fractional charge under $U(1)_Y$,  $U(1)_{Z^\prime}$,
or $U(1)_{em}$.
These exotic states, which do not fit into $SO(10)$ representations,
arise due to the $SO(10)$ symmetry breaking,
by the basis vectors $\alpha$ and $\beta$, and
carry fractional electric charge $\pm1/2$ or
fractional $U(1)_{Z^\prime}$ charge. The full massless spectrum
and charges are given in ref. \cite{fny,cfn}.

In the model of ref. \cite{fny} it is noticed 
that all the fractionally charged states couple at
the cubic level of the superpotential to the set of $SO(10)$
singlets $\{\phi_4,\phi_4^\prime,{\bar\phi}_4,{\bar\phi}_4^\prime\}$ 
\cite{fc},
\beqn
{1\over{\sqrt2}}\{H_1H_2\phi_4
      +(H_3H_4+H_5H_6){\bar\phi}_4
      +(H_7H_8+H_9H_{10})\phi_4'
      +H_{11}H_{13}{\bar\phi}_4'\nonumber\\
\phantom{{0}}
      +(V_{41}V_{42} + V_{43}V_{44}){\bar\phi}_4
      +V_{45}V_{46}\phi_4+(V_{47}V_{48}+V_{49}V_{50}){\bar\phi}_4'
      +V_{51}V_{52}\phi_4'\}.\nonumber
\eeqn
where F--flatness imposes $(\phi_4{\bar\phi}_4'+{\bar\phi}_4\phi_4')=0$.
The problem then is to find flat F and D solutions,
which incorporates the set of fields 
$\{\phi_4,\phi_4^\prime,{\bar\phi}_4,{\bar\phi}_4^\prime\}$.
In the last couple of years the search for F and D flat
solutions in superstring models was systematized \cite{penn}, following
similar developments in supersymmetric field theory models.
Applying these methods to the model of ref. \cite{fny}
we indeed found solutions with the desired properties. 
One example is given by the set of fields, 
$$\{\phi_{12}, \phi_{23}, {\bar\phi}_{56}, 
\phi_4, \phi_4^\prime,
{\bar\phi}_4,{\bar\phi}_4^\prime, H_{15}, H_{30},H_{31}, H_{38} \}.$$
Furthermore, it was shown that with this solution
also the extra Higgs multiplets, beyond the MSSM, as
well as an additional pair of Higgs triplets receive
mass by the same set of VEVs from cubic and
quintic order terms. Therefore, in this solution,
we have in the observable sector, solely the
MSSM charged spectrum. Moreover, the F and D flat solutions
have been completely classified and it was shown that
solutions with such properties are in fact abundant,
which encourages the prospect for obtaining realistic
values for the Standard Model parameters.
Another important property of the F and D flat solutions
is that the set of VEVs necessarily includes fields
that break the $U(1)_{Z^\prime}$, which is embedded in $SO(10)$.
Thus, in this case $SO(10)$ symmetry is necessarily broken directly
to $SU(3)_C\times SU(2)_L\times U(1)_Y$.
Finally, the model of ref. \cite{fny} supplemented
with the flat F and D solutions provides the first example
in the literature with solely the MSSM charged spectrum
below the string scale. Thus, for the first time we
have an example of a long--sought Minimal Superstring Standard Model !

\section{Phenomenology}

The model of ref. \cite{fny} and
its success in the terms of producing solely the MSSM charged
spectrum at low energies, should be viewed as a
prototype example of a realistic free fermionic model.  
The lesson that should be extracted is that
the underlying structure of these models, provided
by the NAHE set, produces the right features for
obtaining realistic phenomenology. It provides further
evidence for the assertion that the true string
vacuum is connected to the $Z_2\times Z_2$ orbifold
in the vicinity of the free fermionic point in the
Narain moduli space. With this in mind we note
that many of the important issues relating to the
phenomenology of the Standard Model and supersymmetric
unification have been addressed in the past
in similar prototype free fermionic heterotic string models. 
These studies have been reviewed in the past and I refer
to the original literature and review references \cite{review}.
These include among others: top quark mass prediction \cite{top}, several
years prior to the actual observation by the CDF/D0 collaborations;
generations mass hierarchy \cite{NRT}; CKM mixing \cite{CKM};
superstring see--saw mechanism \cite{seesaw}; Gauge coupling
unification \cite{gcu}; Proton stability \cite{ps}; and 
supersymmetry breaking and squark degeneracy \cite{fp2}.

\section{Exotic signatures}

After establishing the phenomenological viability of
free fermionic heterotic--string models, it makes 
sense to seek possible experimental signals that
may provide evidence for specific models in particular,
and for string theory in general. This is in essence a secondary task
as the first duty is to reproduce the parameters
of the Standard Model. Obtaining the full structure of
the Standard Model from a string model will be an everlasting
achievement. With this in mind there are several possible
exotic signatures that have been discussed in the past. 
These include the possibility of extra $U(1)$'s \cite{zp};
specific supersymmetric spectrum scenarios \cite{dedes}; R--parity
violation \cite{rp}; and exotic matter \cite{ccf}. 
R--parity violation is
an intriguing but somewhat remote possibility.
The problem is that in string models
if R--parity is violated at the same time one
expects to get fast proton decay.
The model of ref. \cite{custodial} provides an example how
R--parity violation can arise in superstring theory.
This string model gives rise to custodial symmetries
which allow lepton number violation
while forbidding baryon number violation.

The second possibility is that of exotic matter, which
arises in superstring models
because of the breaking of the non--Abelian symmetries
by Wilson--lines. It is therefore a unique
signature of superstring unification, 
which does not arise in field theory GUTs.
While the existence of such states
imposes severe constraints on otherwise valid
string models \cite{otherrsm}, provided that
the exotic states are either confined or sufficiently heavy,
they can give rise to exotic signatures.
For example, they can produce heavy dark
matter candidates, possibly with observable
consequences \cite{ben}.
\section{Toward M(F)--theory embedding}

Over the past few years a remarkable new understanding of
string theory has emerged. In this picture all the ten
dimensional perturbative string theories as well as
11 dimensional supergravity are all perturbative limits
of a single theory, referred to as M (or F) theory \cite{Mtheoryreviews}.
In ref. \cite{befnq} we have undertaken the task
of trying to understand how the phenomenological
free fermionic models may fit in the nonperturbative
framework of M(F)--theory. As discussed above
the NAHE set, which corresponds to $Z_2\times Z_2$
orbifold compactification, plays a pivotal role in
the realistic free fermionic models. Its correspondence
with a $Z_2\times Z_2$ orbifold is made more explicit
by adding to the NAHE set an additional boundary
condition basis vector $X$ \cite{foc}, which extends
the $SO(10)\times SO(6)^3$ symmetry to
$E_6\times U(1)^2\times SO(4)^3$. This model
contains (27,3) multiplets in the $(27,\overline{27})$
representations of $E_6$. The same model is generated
in the orbifold language by moding out an $SO(12)$
Narain lattice by a $Z_2\times Z_2$ discrete symmetry
with standard embedding \cite{foc}. The fermionic
$Z_2\times Z_2$ orbifold model differs from
the one which has usually been examined
in the literature, with $(h_{11},h_{21})=(51,3)$,
which corresponds to twisting of an $SO(4)^3$ lattice.
Many discussions on F--theory have focused on compactifications
on the (51,3) $Z_2\times Z_2$ Calabi--Yau 3--fold to
six dimensions. The first task then is to connect the (51,3) model
to the (27,3) model, and then to implement this connection
in the F--theory compactification on the (51,3) model.
It should be emphasized that the aim here is
not to extract direct phenomenological data from
these investigations. The goal is rather to try
to bridge between basic structures which appear
in the phenomenological free fermionic models
and structures which appear in M(F)--theory, hoping that it will 
eventually yield further phenomenological insight. 

The $(51,3)$ $Z_2\times Z_2$ orbifold model is obtained
by twisting a $(T_2)^3$ 3--dimensional complex manifold
parametrized by $\{z_1,z_2,z_3\}$. The first and second
$Z_2$ twists take $$\{z_1,z_2,z_3\}\rightarrow\{-z_1,-z_2,z_3\}$$
and $$\{z_1,z_2,z_3\}\rightarrow\{z_1,-z_2,-z_3\}.$$
Calculating the cohomology of this manifold,
yields $(h_{11},h_{21})=(51,3)$. Connecting this
model to the $(27,3)$ can be done using the
Landau--Ginzburg formalism with a freely acting
twist \cite{befnq} or by adding the freely acting
shift $$\{z_1,z_2,z_3\}\rightarrow\{z_1+1/2,z_2+1/2,z_3+1/2\},$$
which identifies points by the shift on all three complex
tori simultaneously. Under this shift the number of
fixed points from the twisted sectors is reduced
by half, hence producing the spectrum of the (27,3) model.

The next step is to implement this freely
acting shift or twist in the F--theory compactification
to six dimensions. The key here is that the
models should admit and elliptic fibration with a global section,
in which a Calabi--Yau 3--fold is identified as
a two complex--dimensional base manifold $B$ with a fiber.
The compactification is then defined by specifying the
toroidal fiber in the Weierstrass form, $$y^2=x^3+f(z_1,z_2)x
+g(z_1,z_2),$$ where $f$ and $g$ are polynomials of
degrees 8 and 12, respectively and are functions of
the base coordinates. The number of neutral hypermultiplets,
tensor multiplets and the rank of the vector multiplets
are then given by: $H^0=h^{21}(X)+1$, $T=h^{11}(B)-1$
and $$r(V)=h^{11}(X)-h^{11}(B)-1,$$ respectively, in
terms of $h_{11}$, $h_{21}$ of the CY 3--fold and the base.
Cancelation of the gravitational anomaly in six dimensions
requires that $$H^0-V=273-29T,$$ where $V$ is the number of vector
multiplets. 

The Weierstrass representation of F--theory compactification
on the (51,3) model is given by 
$$y^2 = x^3 + f_8 (w,\tilde w) x z^4 + g_{12}(w,\tilde w) z^6,$$
where $$f_8=\eta-3h^2,~{\rm and}~g_{12}~=~h(\eta-2 h^2),$$
$$h = K \prod_{i,j=1}^4(w- w_i)(\tilde w - \tilde w_j)$$ and
$$\eta = C \prod_{i,j=1}^4(w- w_i)^2(\tilde w - \tilde w_j)^2.$$
Taking $w\rightarrow w_i$ (or ${\tilde w}\rightarrow
{\tilde w}_i$) we have a $D_4$ singular fiber. Thus, we
have an enhanced $SO(8)^8$ gauge symmetry, since $i,j=1,\ldots,4$.
These $D_4$ singularities intersect in 16 points, $(w_i,\tilde w_j),\,
i,j=1,\ldots 4$, in the base and give rise to 16 additional
tensor multiplets. With the equations given above we see that
this fits the data of F--theory compactification on the
$Z_2\times Z_2$ CY 3--fold with $(h_{11},h_{21})=(51,3)$.

To find the F--theory compactification on the corresponding
$(27,3)$ model we implement the freely acting twist in the
elliptic fibration. For this purpose it is more convenient
to represent the fiber in quartic form, which is given by
$$\hat y^2 = \hat x^4 + \hat x^2 \hat z^2 \hat f_4 + \hat x z^3 \hat g_6
+ \hat z^4 \hat h_8,$$ with $\hat f_4 = -3 h$,
$\hat g_6 = 0$, and $\hat h_8 = -1/4 \eta.$
The freely acting twist which acts simultaneously
on the base and the fiber is given by
$$(\hat y,\hat x, \hat z,w,\tilde w)\to (-\hat y,-\hat x, \hat
z,-w,-\tilde w).$$ We now see that to implement this
identification $h$ and $\eta$ are modified and are given by
$$h = K \prod_{i,j=1}^2(w^2- w_i^2)(\tilde w^2 - \tilde w_j^2)$$ and 
$$\eta = C \prod_{i,j=1}^2(w^2- w_i^2)^2(\tilde w^2 - \tilde w_j^2)^2.$$
We see that there are now only 4 $D_4$ singularities, and
similarly the number of intersecting singularities is reduced
by 2. Hence, in this F--theory compactification the enhanced
symmetry is $SO(8)^4$ with $T=9$, $H^0=4$ and $V=112$.
This matches the data of the (27,3) model with $h_{11}(B)=10$.
However, we see that in this case the gravitational
anomaly is apparently not satisfied.

A plausible interpretation of the above result is that
the (27,3) model does not provide a consistent background
for F--theory compactification to six dimension. However,
this will still be a strange situation because the freely acting twist
is a consistent operation which should not destroy the fibration.
To study the issue further, we examine the effect of
the freely acting shift on the $(T_2)^3$ manifold.
It is then seen that although the shift is freely
acting on the CY 3--fold, it is not freely acting on the
base when the CY is regarded as a fibration.
Hence, the base in the fibered CY has four singular
points that are not singular points of the CY 3--fold.
Therefore, there are no $h_{11}$ and $h_{21}$ forms
that can be used to resolve these singularities.
The elliptic fibration and the global section
are destroyed at those singular points. 

The existence of these special singular points is
quite interesting. Another plausible interpretation
for the resolution of the puzzles is that due
to these special singularities there exist
additional massless states that can only be
seen nonperturbatively. Furthermore, the appearance
of the special singularities is closely tied
to the action of the freely acting shift on
the fiber. To see that we implement another
shift on the elliptically fibered (51,3) model,
induced by the shift $$(z_1,z_2,z_3)\rightarrow
(z_1+{1/2},z_2+{1/2},z_3).$$ This shift
is not freely acting on the CY 3--fold and there is
an additional sector yielding $(h_{11},h_{21})=(31,7)$.
This shift is not freely acting on the base
but there are now four additional $(h_{11},h_{21})$
pairs that can be used to resolve the base singularities
in the usual manner. For F--theory compactification
on the (31,7) CY 3--fold: $T=13;~H^0=8$ and $V=112$
and it is checked that it is consistent with
the gravitational anomaly. 

The discussion above illustrates that the puzzles
in the F--theory compactification on the (27,3)
model precisely arise because of the action of
the freely acting shift on the fiber
$$z_3\rightarrow z_3+1/2.$$ While one possible
interpretation is that the (27,3) model
does not provide a consistent background for
F--theory compactification, the situation is
non trivial and still unresolved. But we may contemplate
that some new physics is associated with the action
of the shift on the fiber, which in the case
of the type IIB string theory is identified
with the dilaton. All in all we see that the
free fermionic $Z_2\times Z_2$ orbifold is
special and perhaps more surprises lie in
store. 
\section{Equivalence postulate of quantum mechanics --\\
$~~~~~~~~~~$embarking on a novel approach to quantum gravity}
Over the last few years important new insight has been gained
on the fundamental structure of string theory. We now know
that different theories, that are classically distinct,
are in fact related quantum mechanically
by various duality transformations. Many of the duality
relations have an inherent geometrical description. 
What is the lesson to be extracted from this new
understanding ? It seems to me that 
what is needed is a new fundamental
principle. This is the approach which
is being pursued by Matone and myself. We have imposed
the basic postulate that all physical systems labelled
by a potential function can be connected by a coordinate
transformation and showed that consistency of
this postulate necessitates the appearance of
quantum mechanics and is intimately connected
to phase--space duality. 
The Planck constant appears in this context
as a covariantizing parameter. We then have a
fundamental geometrical principle behind
quantum mechanics and $\hbar\ne0$. I will
follow here the historical path of this development. 

Working in Seiberg--Witten theory \cite{SW} Matone \cite{inverrel} noted
that the Picard--Fuchs equation for the duals, $a(u)$ and 
$a_D(u)$, can be inverted. Yielding
$u(a)={1\over2} a {\partial_a} {\cal F}-{\cal F}$,
which is an exact non--perturbative relation
and the prepotential function is given by
$a_D={\partial_a{\cal F}}$. Written in the form
$u(a)=a^2{\partial_{a^2}}{\cal F}-{\cal F}$,
it is noted that it has the form of the Legendre
transformation. This relation is not particular
to Seiberg--Witten theory and can be applied to other
theories which are described by a second order linear
differential equation. We first applied this idea
to the Schr\"odinger equation, where we introduced
a prepotential function defined by the relation
$$\psi_D=\partial_\psi{\cal F},$$ with $\psi$ and $\psi_D$
being the two linearly independent solutions of
the Schr\"odinger equation \cite{fm1}. One then obtains the
inverted form $$x(\psi)=\psi^2\partial_{\psi^2}{\cal F}-{\cal F},$$
which offers the possibility of a coordinate free
formulation of quantum gravity. A direction which
is being pursued primarily in ref. \cite{vancea,carroll}. 

The inversion relation is a general relation between
dual variables related by a generating function. The
natural step is to apply it to the phase--space
coordinates related by Hamilton's generating function
$p=\partial_q{\cal S}_0$. One then obtains the 
dual Legendre transformations \cite{fm2},
$${\cal S}_0=p\partial_p{\cal T}_0-{\cal T}_0$$ and
$${\cal T}_0=q\partial_q{\cal S}_0-{\cal S}_0$$
where ${\cal T}_0(p)$ is a new generating function
defined by $q=\partial_p{\cal T}_0$.
Two observations are important to note.
The first is that the Legendre transformation
is undefined for
linear functions, {\it i.e.} for ${\cal S}_0=A+Bq$. 
The second is that similar to the case
of Seiberg--Witten theory, and the case
of the Schr\"odinger equation, one can 
associate a second order differential
equation with each Legendre transformation,
which we call the ``canonical equation'' \cite{fm2}.
The potential function in the ``canonical
equation'' for ${\cal S}_0$ is given by
$\{q,{\cal S}_0\}$, where $\{,\}$ denotes
the Schwarzian derivative. Choosing that the
reduced action transforms as a scalar function
under coordinate transformations, it is noted that 
by construction the $2^{nd}$--order
differential equation is covariant. The Schwarzian
derivative, however, is invariant under M\"obius
transformations, but not under general
coordinate transformations.
This fact suggests that different physical
systems labelled by different potentials can be connected
by coordinate transformations.
Given the new insight gained in the context
of string dualities, and the discussion above
on Legendre duality, it is natural to promote
this new insight to the level of a fundamental physical principle.
In ref. \cite{fm2} we posed the following postulate:

{\it Given two physical systems with ${\cal W}^a(q^a)\in{\cal H}$
and ${\cal W}^b(q^b)\in{\cal H}$, where ${\cal H}$ denotes
the space of all possible $\cal W$'s, there always exists a 
coordinate transformation $q^a\rightarrow q^b=v(q^a)$ 
such that ${\cal W}^a(q^a)\rightarrow {\cal W}^{av}(q^b)={\cal W}^b(q^b)$.}

We note that this postulate also implies that there should
always exist a coordinate transformation connecting
any state  to the state ${\cal W}^0(q^0)=0$. Inversely, this means
that any state ${\cal W}\in{\cal H}$ can be reached from the
state ${\cal W}^0(q^0)$ by a coordinate transformation. 

A natural application of this postulate is in the context
of the classical Hamilton--Jacobi formalism. There
one solves the dynamical problem by performing canonical
transformations which map a dynamical system, governed
by a Hamiltonian $H$, to a trivial dynamical system
with vanishing Hamiltonian. The solution is given
by the Classical Hamilton--Jacobi Equation (CHJE),
and the functional relation between $p$ and $q$
is only extracted after the Hamilton--Jacobi equation
is solved.
We aim to pose a similar question, but with the
novelty that we consider the transformation $q\rightarrow{\tilde q}(q)$,
while imposing the functional relation $p=\partial_q{\cal S}_0$,
reducing to the free system with vanishing energy.
Motivated from the Legendre duality discussion
we impose that under the transformation
${\tilde {\cal S}}_0({\tilde q})={\cal S}_0(q)$.
It follows that $p$ transforms as $\partial_q$.

The CSHJE, $${1/{2m}}({\partial_q{\cal S}_0})^2+{\cal W}(q)=0,$$
fixes the transformation
$${\cal W}(q)\rightarrow {\tilde {\cal W}}({\tilde q})=
(\partial_{\tilde q}q)^2{\cal W}(q).$$
It is observed that the state ${\cal W}^0(q^0)=0$
is a fixed point under the coordinate transformation.
That is we cannot reach all possible states by coordinate
transformation from the state ${\cal W}^0(q^0)=0$.
Consistency of the equivalence postulate then implies that
the CSHJE should be deformed. The most general form would be,
$${1/{2m}}({\partial_q{\cal S}_0})^2+{\cal W}(q)+{Q}(q)=0,$$
where the nature of ${Q}(q)$ is to be determined by
the consistency of the equivalence postulate,
which imposes that the combination $({\cal W}+{Q})$
transforms as a quadratic differential. On the other
hand all states should be connected to the state
${\cal W}^0(q^0)=0$ by a coordinate transformation.
The basic transformation properties are,
\beqn
{\cal W}^v(q^v)&=&
 \left(\partial_{q^v}q^a\right)^2{\cal W}^a(q^a)+(q^a;q^v),\nonumber\\
 Q^v(q^v)&=&\left(\partial_{q^v}q^a\right)^2Q^a(q^a)-(q^a;q^v),\nonumber
\eeqn
which fixes the cocycle condition for the inhomogeneous term \cite{fm2}
$$
(q^a;q^c)=\left(\partial_{q^c}q^b\right)^2[(q^a;q^b)-(q^c;q^b)].
$$
The importance of the cocycle condition is that it uniquely
fixes the transformation properties of the inhomogeneous
term, and hence fixes its functional form. It is then
proven that the inhomogeneous term $(q^a;q^b)$ is
invariant under the M\"obius transformation, and is
uniquely given by the Schwarzian derivative $\{q^a,q^b\}$.
The cocycle condition is generalizable
to higher dimensions and fixes that the inhomogeneous
term is invariant under the D--dimensional M\"obius transformations
\cite{bfm}.
The cocycle condition univocally implies \cite{fm2}, 
\beqn
&&{\cal W}(q)= V(q) - E = -{\hbar^2\over{4m}}\{
{\rm e}^{(i2{\cal S}_0/\hbar)},q\},
\label{wschwarz}\\
&&{Q}(q)~~~~~~~~~~~~~~~~~=  {\hbar^2\over{4m}}\{{\cal S}_0,q\},
\label{qschwarz}
\eeqn
where in demonstrating this we used the basic identity,
\beq
({\partial_q{\cal S}_0})^2=
\hbar^2/2\left(\{\exp(i2{\cal S}_0/\hbar,q)\}-\{{\cal S}_0,q\}\right).
\label{qid}
\eeq
${\cal S}_0$ is solution of the Quantum Stationary
Hamilton--Jacobi Equation
\beq
{1\over{2m}}\left({{\partial S}_0\over{\partial q}}\right)^2+
V(q)-E+{\hbar^2\over{4m}}\{{\cal S}_0,q\}=0,
\label{qhje}
\eeq
which can be obtained from the
Schr\"odinger equation by taking,
$$\psi(q)={1\over{\sqrt{{\cal S}_0^\prime}}}
{\rm e}^{\pm{{i{\cal S}_0\over\hbar}}}$$
Note that the QSHJE is a non--linear third--order differential
equation. The Schr\"odinger equation in this context can
be regarded as linearization of the QHJE
\cite{fm2}, in the following sense.
From eq. (\ref{wschwarz})
and the M\"obius invariance of the Schwarzian derivative
it is seen that the solution of the QHJE is given in terms 
of the ratio of the two real linearly independent solutions of 
the stationary Schr\"odinger equation, $w=\psi_D/\psi$, by
\beq
{\rm e}^{{{i2}\over\hbar}{\cal S}_0\{\delta\}}=
{\rm e}^{i\alpha}{{w+i{\bar\ell}}\over{w-i\ell}}
\label{s0intermsofw}
\eeq
where $\delta=\{\alpha,\ell\}$ with $\alpha\in R$ and ${\rm Re}\ell\ne0$,
which is equivalent to the condition ${\cal S}_0\ne const$.
We note that the trivializing map to the ${\cal W}^0(q^0)=0$
system is given by $q\rightarrow q^0=w$.

Several points are
important to note. First is that also for the state
${\cal W}^0=0$ we have ${\cal S}_0\ne const$.
Therefore, ${\cal S}_0=const$ is not in the space of
solutions and we have that the equivalence postulate is consistent
with quantum mechanics but is inconsistent with classical mechanics.
This fact also allows for the definability of
the Legendre transformation for all physical states.
Thus we have that the definability of the Legendre
duality and the consistency of the equivalence postulate
are intimately related.

It is further shown \cite{fm2} that consistency of the equivalence
postulate implies both energy quantization 
for bound states with a square integrable wave--function,
as well as the tunnelling effect, without assuming the
probability interpretation of the wave--function.
Thus, we have that the main characteristics of quantum
mechanics arise from the self--consistency of the
equivalence postulate. This is of fundamental importance
as we see that the main phenomenological features
of quantum mechanics are reproduced starting
from the equivalence postulate
without further assumptions. I refer the reader
to the original papers \cite{fm2} for details of this
fascinating avenue. Here I focus on the characteristics
of the formulation which are related to the Planck 
scale, and hence may be related to string theory.

There are two consequences of the formulation that are
clearly related to the Planck scale and hence to gravity and
possibly to string theory. The first is the appearance
of a fundamental length scale which is identified with the
Planck length and the second is the existence of equivalence
classes of the wave--function which depend on this
length scale \cite{fm2}.

First it is noted that the formulation provides a trajectory
representation of quantum mechanics \cite{floyd},
which due to the M\"obius symmetry depends on the constant
$\ell$. From the solution for the
QHJE we get $p={\partial_q {\cal S}_0}= p(q)$, which
depends on the integration constants of the QSHJE.
\begin{equation}
p_E=\pm{\hbar (\ell_E+\bar\ell_E)\over 2|k^{-1}\sin kq-i\ell_E \cos kq|^2},
\label{dajjj}
\end{equation}
where $k=\sqrt{2mE}/\hbar$. 
The existence of a fundamental length scale, identified
with the Planck length, can already be inferred from the
basic Legendre duality and consistency of the equivalence
postulate which require that ${\cal S}_0\ne0$. 
It is seen more explicitly by considering the
consistency of the 
classical limit, $\hbar\rightarrow 0$. 
In this limit we have that for $E\rightarrow0$ 
we should have $p_0\rightarrow0$.
This shows that Re$\ell_0\sim \hbar^\gamma$
with $-1<\gamma<1$. Thus, consistency of these limits
implies the identification 
${\rm Re}\ell_0\sim\lambda_P\ne0$ \cite{fm2}.

The next important property of the formulation is
the existence of equivalence classes of the wave--function.
As the QHJE is a third--order differential equation whereas the Schr\"odinger
equation is a second order one, more initial conditions
are needed to be specified in the case of the QHJE.
It follows that the wave function remains
invariant under suitable transformations of $\delta=\{\alpha, \ell\}$,
corresponding to different trajectories.
The implication is that there
are hidden variables which depend of the Planck length and
that these can suitably change without affecting the 
wave--function. Recently, t'Hooft \cite{thooft} has
advocated that hidden variables must play a role
in the implementation of the holographic principle \cite{holo}.
\section{Is there a connection with string theory ?}
At the outset I would state that I do not know
the answer to this question. 
The aim is to try to find some
overlaps in the physical and mathematical characteristics
in the equivalence postulate derivation and in string theory.
Nevertheless, it should be stressed that it is very natural
to expect that the correct theory of quantum gravity 
would arise from a principle such as the equivalence postulate.
Already the appearance of the Schwarzian derivative
in the framework of quantum mechanics should be
regarded as tantalizing evidence for a possible
connection with string theory and hence with quantum gravity.
Below I enumerate other possible relations.

1) Quadratic differential: In string theory elimination
of the world--sheet conformal anomaly is necessary in
order for the energy--momentum tensor to transform 
as a quadratic differential and for obtaining
diffeomorphism invariance, {\it i.e.} Einstein
equations, in target space. Thus, the fact that the
energy--momentum tensor transforms as a quadratic
differential plays an important role. Similarly,
the equivalence postulate derivation imposes that the
Hamiltonian (Hamilton--Jacobi equation)
transforms as a quadratic differential. The similarity
is not complete because in the string case we
require that cancelation of a quantum anomaly
restores a classical symmetry, whereas in the
equivalence postulate derivation we required that the
quantum modification enables that the HJ equation
transforms as a quadratic differential. Nevertheless,
it seems that there should be a deep reason
why in both cases the quadratic differential transformation
plays a crucial role. Another caveat is that
we have not yet included fermions in the formalism.
However, we may envision that the square of the fermionic
Hamiltonian transforms as a quadratic differential.

2) The existence of M\"obius symmetry represents
an invariance under finite diffeomorphism. The 
$SL(2,C)$ symmetry plays a central role in string theory
and a central role in the formulation of quantum
mechanics from the equivalence postulate.
The M\"obius symmetry is the origin of the
existence of equivalence classes of the
wave--function and of a fundamental length scale.
The presence of a M\"obius symmetry should suggest
a connection with string theory.
Indeed, one may expect that performing
infinitesimal diffeomorphism would recover some 
of the features of 2D-CFT's, including
the Virasoro algebra, vector spaces, etc.
In higher dimensions the symmetry of the cocycle
is the higher dimensional M\"obius group \cite{bfm}. In that case,
we may envision that performing infinitesimal diffeomorphism
should entail some generalization of the Virasoro
algebra, possibly in the form of W--algebras ?

3) The hidden variables in the
equivalence postulate formulation can be identified
with a fundamental length scale, most naturally with
the Planck length \cite{fm2}. Furthermore, the equivalence
classes of the wave--function can be parametrized
in terms of this fundamental length scale.
The emergence of a fundamental length scale
should give rise to the suspicion of
a connection with quantum gravity and
possibly with string theory. An intriguing thought
is that the emerging length scale and its role
in the equivalence classes of the wave--function
may somehow be related to the internal string dimension.
The analogy of the formalism with uniformization
theory \cite{fm2} suggests that the hidden variables may be associated
with Riemann surfaces, further indicating possible connections
with string theory.
\section{Does it address the vacuum energy problem ?}
Again I would state that I do not know the answer
to that question. We may however contemplate how
the equivalence postulate may affect the standard
picture. Surely, if the equivalence postulate
is a fundamental law of Nature, as we may infer
from the understandings gained in the context
of string dualities, then it will by default
also shed light on this issue. 

We see that in the quantum HJ equation
there is an additional term which is identified
with a quantum motion, or quantum potential,
which is nonzero also when the potential and the
energy vanish
\begin{equation}
{1\over{2m}}\left({\partial_q{\cal S}_0}\right)^2+{\hbar^2\over{4m}}
\{{\cal S}_0,q\}=0
\label{hjvpve}
\end{equation}
The implication is that, unlike the classical case, ${\cal S}_0$
is never vanishing and the state ${\cal S}_0=constant$ is excluded
from the space of allowed solutions. This is the fundamental
characteristic of quantum mechanics in our approach.
Thus, we see that the state with $V(q)=0$ and $E=0$
is indeed pointed out in the equivalence postulate
approach. The solution for the ground state
in the quantum case is given by ${\cal S}_0={i\hbar/2}\ln{q}$,
up to M\"obius transformations. Consequently,
the conjugate momentum for the ground state
is also non--trivial. Being the characteristic property
of quantum mechanics, we can regard this as the
quantum trajectory version of the uncertainty principle.
All in all, we see that the vacuum state is singled
out relative to its role in the classical case.
Furthermore, the quantum ground solution is also
the self--dual state of the Legendre phase--space
transform and its dual. That is, it is the unique
simultaneous solution of the two second order linear differential
equations associated with each Legendre transformations \cite{fm2}.
Thus, we have that the vanishing of the vacuum state
may be intimately related to the Legendre phase--space duality. 
This is already one hint that the equivalence
postulate and Legendre phase--space duality
may shed light on the vacuum energy issue.

The equivalence postulate point of view is that
the fundamental equation is the QHJE, which is
a third--order non--linear differential equation.
It is equivalent to the Schr\"odinger equation (in the sense discussed
above eq. (\ref{s0intermsofw})),
but requires specifying more initial conditions
than for the Schr\"odinger equation. We
have that there is a moduli space of solutions
of the QHJE, which corresponds to the same
wave function. That is, there are hidden
variables which depend on the Planck length and
are not detected in the solutions of the
Schr\"odinger equation. This means that the
Schr\"odinger equation with its related
apparatus provides an effective description,
albeit an extremely successful one from
the experimental point of view. Now, the vacuum
energy in conventional quantum mechanics is an
artifact of the Hilbert space construction, {\it i.e.}
it is an artifact of the effective description.
But from the
point of view of the equivalence postulate
the more complete solution is given by the QHJE,
which admits a non--trivial solution also for the
state with vanishing energy and vanishing potential.
The existence of such a specialized state
already indicates that it may have something
to do with the vacuum energy, as according to
the equivalence postulate all other states
are connected to this special state by
coordinate transformations. This leads
to the existence of a fundamental length
scale with all the expected implications
of modifications of the uncertainty relations,
and space--time uncertainty relations, etc.
However, the important fact is the existence
of the additional term in the quantum HJ equation,
$Q(q)=(\hbar/2m)\{{\cal S}_0,q\}$, which is never
vanishing. This term can be interpreted as
a curvature term \cite{fm2}, which means that the existence
of the state itself is associated with a sort of ``quantum
curvature''. So it seems artificial to speak of
particles and curvature as distinct entities.
It seems, that from the equivalence postulate
point of view there is no meaning to talk
about vacuum energy. Rather we should speak about the vacuum
curvature, associated with the existence
of the particles themselves, which is never vanishing,
and is associated with the state for which the
potential and the energy vanish.

Another observation on the equivalence postulate derivation
is the way in which mass appears in the formalism.
In quantum field theories, the
vacuum energy problem is tightly related to the generation
of mass through symmetry breaking. 
In the equivalence postulate derivation, it is intriguing to note
that mass appears only after making the identification
in eq. (\ref{wschwarz}). In the identity, eq. (\ref{qid}),
which is related to the QHJE eq. (\ref{qhje}), the mass
is only a multiplicative constant, which can be dropped from
the equation. This feature also persists in the higher
dimensional generalization \cite{bfm}. Thus, we see that mass in
the equivalence postulate derivation is an artificial
effect of representing the potential in the functional
form ${\cal W}(q)=V(q)-E$. The fact that the Schwarzian
derivative is associated with a curvature term suggests
that mass in this formalism is related to intrinsic
curvature associated with a particle state.
This property of the formalism
is reminiscent of the gravitational equivalence principle 
and is further evidence that the equivalence postulate
formalism provides a natural framework for quantum gravity.
\section{Conclusions}
Should we believe in the relevance of string theory in nature ? 
The most urgent issue in particle physics is the nature
of the electroweak symmetry breaking mechanism.
The basic question is whether fundamental scalar states
exist in nature, or whether a more intricate, yet
unperceived, mechanism plays a role. This question must and will be
resolved by our experimental colleagues, who have already provided
us with a glorious confirmation of the Standard Model
gauge and matter sectors.
The Standard Model multiplet structure strongly indicates
the realization of grand unifying structures in nature,
and augmented with supersymmetry is the leading candidate
for a theory of electroweak symmetry breaking, to be tested
by future experiments. String theory provides the most advanced
tools to study quantum, gravity and gauge, unification.
The fact that one finds string models that closely resemble
the real world, and exactly where expected,
namely near a maximally symmetric point, leads to the intriguing
suspicion that string theory is indeed relevant in nature. 

A plausible view of recent years deeper understandings
in string theory is that different theories that are
classically distinct, are in fact connected quantum
mechanically. It is then natural to promote this new
insight to a level of a fundamental physical principle.
From such a principle the correct theory of
quantum gravity should be derivable, as well as 
the phenomenological characteristics of quantum
field theories, and the fundamental vexing problems,
like the vanishing of the cosmological constant and
the problem of mass. At the closing of one millennium,
it seems that the new one may still offer plenty of
surprises. 


\section*{Acknowledgments}
I would like to thank Per Berglund, Gaetano Bertoldi, Jerry Cleaver,
John Ellis, Dimitri Nanopoulos, Keith Olive, Maxim Pospelov, Zongan
Qiu and especially Marco Matone, for collaboration on part of the work
reported in this paper, and Marco Matone and Adam Ritz for comments on
the manuscript. I would like to thank the CERN, ITP--UCSB and University
of Padova theory groups for hospitality. This work is supported in part
by DOE grant No. DE--FG--0287ER40328.
\bibliographystyle{unsrt}

\begin{thebibliography}{99}
\bibitem{gg}      Georgi H and Glashow S 1974 \PRL {\bf32} {438};
                    H. Georgi, in {\it Particles and Fields--1974},
                    ed. C.E. Carlson. 1975, New York, AIP Press.
\bibitem{lds} Krauss L M and Wilczek F \PRL{62}{89}{1221};
              Faraggi A E \PLB{398}{97}{88}.
\bibitem{gqw} Georgi H \etal 1974 \PRL {\bf33} {451}.
\bibitem{gcumssm} Dimopoulos \etal \PRD{24}{81}{1681};
                  Ellis J \etal \PLB{260}{91}{131};
                  Langacker P and Luo M \PRD{44}{91}{817};
                  Amaldi U\etal \PLB{260}{91}{447}.
\bibitem{stringguts} Lewellen D C \NPB{337}{90}{61};
     Aldazabal G \etal \NPB{452}{95}{3};
     Cleaver G hep-th/9506006;
     Kakushadze Z and Tye S H H 1996 \PRL {\bf77} 2612;
                                        \PRD{55}{97}{7896};
                                        hep-ph/9705202.
\bibitem{suthree} Candelas P \etal
                                        \NPB{258}{85}{46};
                Greene B \etal
                                        \NPB{292}{87}{606};
                Arnowitt R and  Nath P 1989 \PRL {\bf62} {2225};
\bibitem{revamp} {Antoniadis I \etal \PLB{231}{89}{65}.}
\bibitem{patisalamstrings} Antoniadis I \etal
                                        \PLB{245}{90}{161};
                Leontaris G K and Rizos J hep-ph/9901098.
\bibitem{zthree} Font A \etal \NPB{331}{90}{421};
                Casas J A \etal\NPB{317}{89}{171}.
\bibitem{fny} {\AEF \etal \NPB{335}{90}{347}.}
\bibitem{eu} \AEF \PLB{278}{92}{131}; \NPB{387}{92}{239}.
\bibitem{top} {\AEF \PLB{274}{92}{47}.}
\bibitem{otherrsm} Chaudhuri S \etal \NPB{469}{96}{357};
\bibitem{ps} Faraggi A E \NPB{428}{94}{111}; Pati J C, \PLB{388}{96}{532}
Ellis J \etal \PLB{419}{98}{123}.
\bibitem{twozero} Witten E \NPB{268}{86}{79}; 
                  Distler J and Kachru S \NPB{413}{94}{213}.
\bibitem{dhvw} Dixon L \etal \NPB{274}{86}{285}.
\bibitem{narain} Narain K S \PLB{169}{86}{41};
                 Narain K S \etal \NPB{279}{87}{369}.
\bibitem{foc} \AEF \PLB{326}{94}{62}.
\bibitem{FFF} Kawai K \etal \NPB{288}{87}{1};
                Antoniadis I \etal \NPB{289}{87}{87}.
\bibitem{KLN} Dixon L \etal \NPB{282}{87}{13};
              Kalara S \etal \NPB{353}{91}{650}.
\bibitem{nahe} Faraggi A E and Nanopoulos D V \PRD{48}{93}{3288};
               Faraggi A E hep-th/9511093; hep-th/9708112.
\bibitem{dsw}
         Dine M \etal \NPB{289}{87}{589};
         Atick J J \etal \NPB{292}{87}{109};
         Cecotti S \etal \IJMP{2}{87}{1839}.
\bibitem{cfn} Cleaver G \etal \PLB{455}{99}{135}; hep-ph/9904301; 
hep-ph/9910230.
\bibitem{fc} Faraggi A E \PRD{46}{92}{3204}.
\bibitem{penn} Cleaver G \etal \NPB{525}{98}{3}; \NPB{545}{98}{47};
                                                \PRD{59}{99}{055005};
                                                \PRD{59}{99}{115003}.
\bibitem{review} For review see {\it e.g.} Lykken J hep-ph/9511456;
Lopez J L hep-ph/9601208; Faraggi A E hep-ph/9404210; hep-ph/9707311.
\bibitem{NRT} \AEF \NPB{403}{93}{101};\NPB{407}{93}{57}.
\bibitem{CKM} Antoniadis I \etal \PLB{278}{92}{257};
\AEF and Halyo E \PLB{307}{93}{305}; \NPB{416}{94}{63};
Ellis J \etal \PLB{425}{98}{86}.
\bibitem{seesaw} Antoniadis I \etal \PLB{279}{92}{281};
             Faraggi A E and Halyo E \PLB{307}{93}{311};
             Faraggi A E and Pati J C \PLB{400}{97}{314}. 
\bibitem{gcu} Antoniadis I \etal \PLB{268}{91}{188}; 
              Antoniadis I \etal \PLB{272}{91}{31};
              Faraggi A E \PLB{302}{93}{202}; 
              Dienes K R and Faraggi A E 1995 \PRL {\bf75} {2646};
                                           \NPB{457}{95}{409}.             
\bibitem{fp2} Antoniadis I \etal \PLB{241}{90}{24}; 
              \AEF and Halyo E \IJMP{11}{96}{2357};
              Faraggi A E and Pati J C \NPB{526}{98}{21}.
\bibitem{zp} Costa G \etal \NPB{297}{88}{244};
             Hewett J and Rizzo T \PRT{183}{89}{193};
             Faraggi A E and Nanopoulos D V \MODA{6}{91}{61};
             Cvetic M and Langacker P hep-ph/9707451.
\bibitem{dedes} Dedes A and Faraggi A E hep-ph/9907331.
\bibitem{rp} Faraggi A E \PLB{398}{97}{95}.
\bibitem{ccf} Chang S \etal \PLB{397}{97}{76}; \NPB{477}{96}{65}; 
              Elwood J and \AEF \NPB{512}{98}{42}.
\bibitem{custodial} \AEF \PLB{339}{94}{223}.
\bibitem{ben} Benakli K \etal \PRD{59}{99}{047301}; 
             Faraggi A E \etal hep-ph/9906345.
\bibitem{Mtheoryreviews} Schwarz J hep-th/9607201;
                         Townsend P K hep-th/9612121;
                         Vafa C hep-th/9702201;
                         Sen A hep-th/9802051;                  
                         Duff M J hep-th/9805177;
                         Li M hep-th/9811019.
\bibitem{vafamorrison} Morrison D R and Vafa C \NPB{473}{96}{74};
                                               \NPB{476}{96}{437}.
\bibitem{befnq} Berglund P \etal \PLB{433}{98}{269}; hep-th/9812141.
\bibitem{SW} Seiberg N and Witten E, \NPB{426}{94}{19}.
\bibitem{inverrel} Matone M \PLB{357}{95}{342}.
\bibitem{fm1} \AEF and Matone M 1997 \PRL {\bf78} 163.
\bibitem{vancea} Vancea I V gr-qc/9801072; 
                  De Andrade M A and Vancea I V gr-qc/9907059.
\bibitem{carroll} Carroll R hep--th/9607219; hep-th/9610216;
hep-th/9702138; hep-th/9705229; quant-ph/9903081.
\bibitem{fm2} \AEF and Matone M \PLB{450}{99}{34}; \PLB{437}{98}{369};
\PLA{249}{98}{180}; \PLB{445}{98}{77}; \PLB{445}{99}{357}; hep-th/9809127.    
\bibitem{floyd} Floyd E R \PRD{25}{82}{1547}; \PRD{26}{82}{1339};
\PRD{29}{84}{1842}; \PRD{34}{86}{3246}; \PLA{214}{96}{259}; 
\IJMP{14}{99}{1111}.
\bibitem{bfm} Bertoldi G \etal, hep-th/9909201.
\bibitem{thooft} 't Hooft G gr-qc/9903084;
\bibitem{holo} 't Hooft G gr-qc/9310026; Susskind L hep-th/9409089.
\end{thebibliography}

\vfill\eject
\end{document}